\documentclass[8.5pt,twoside,twocolumn]{article}
\usepackage{ulem} 
\usepackage[super,sort&compress,comma]{natbib} 
\usepackage{mhchem}
\usepackage{times,mathptmx}
\usepackage{sectsty}
\usepackage{balance} 
\usepackage{graphicx}
\usepackage{lastpage}
\usepackage[format=plain,justification=raggedright,singlelinecheck=false,font=small,labelfont=bf,labelsep=space]{caption} 
\usepackage{fancyhdr}
\usepackage{url}
\usepackage{color}
\usepackage{wasysym}

\begin{document}
%\thispagestyle{plain}
%\fancypagestyle{plain}{
%\fancyhead[L]{\includegraphics[height=8pt]{headers/LH}}
%\fancyhead[C]{\hspace{-1cm}\includegraphics[height=20pt]{headers/CH}}
%\fancyhead[R]{\includegraphics[height=10pt]{headers/RH}\vspace{-0.2cm}}
%\renewcommand{\headrulewidth}{1pt}}
\renewcommand{\thefootnote}{\fnsymbol{footnote}}
%\renewcommand\footnoterule{\vspace*{1pt}
%\hrule width 3.4in height 0.4pt \vspace*{5pt}} 
%\setcounter{secnumdepth}{5}
%\makeatletter 
%\def\subsubsection{\@startsection{subsubsection}{3}{10pt}{-1.25ex plus -1ex minus -.1ex}{0ex plus 0ex}{\normalsize\bf}} 
%\def\paragraph{\@startsection{paragraph}{4}{10pt}{-1.25ex plus -1ex minus -.1ex}{0ex plus 0ex}{\normalsize\textit}} 
%\renewcommand\@biblabel[1]{#1}            
%\renewcommand\@makefntext[1]
%{\noindent\makebox[0pt][r]{\@thefnmark\,}#1}
%\makeatother 
%\renewcommand{\figurename}{\small{Fig.}~}
%\sectionfont{\large}
%\subsectionfont{\normalsize} 
%\fancyfoot{}
%\fancyfoot[LO,RE]{\vspace{-7pt}\includegraphics[height=9pt]{headers/LF}}
%\fancyfoot[CO]{\vspace{-7.2pt}\hspace{12.2cm}\includegraphics{headers/RF}}
%\fancyfoot[CE]{\vspace{-7.5pt}\hspace{-13.5cm}\includegraphics{headers/RF}}
%\fancyfoot[RO]{\footnotesize{\sffamily{1--\pageref{LastPage} ~\textbar  \hspace{2pt}\thepage}}}
%\fancyfoot[LE]{\footnotesize{\sffamily{\thepage~\textbar\hspace{3.45cm} 1--\pageref{LastPage}}}}
\fancyhead{}
\renewcommand{\headrulewidth}{1pt} 
\renewcommand{\footrulewidth}{1pt}
\setlength{\arrayrulewidth}{1pt}
\setlength{\columnsep}{6.5mm}
\setlength\bibsep{1pt}
\twocolumn[
\begin{@twocolumnfalse}
\noindent\LARGE{\textbf{Approach to universal self-similar attractor for the levelling of thin liquid films}}
\vspace{0.6cm}

\noindent\large{\textbf{Michael Benzaquen\textit{$^{a*}$}, Paul Fowler\textit{$^{b*}$}, Laetitia Jubin\textit{$^{b}$}, Thomas Salez\textit{$^{a}$}, Kari Dalnoki-Veress\textit{$^{a,b}$} and Elie Rapha\"el$^{\ddagger}$\textit{$^{a}$}}}\vspace{0.5cm}

%\noindent\textit{\small{\textbf{Received Xth XXXXXXXXXX 20XX, Accepted Xth XXXXXXXXX 20XX\newline
%First published on the web Xth XXXXXXXXXX 200X}}}
%
%\noindent \textbf{\small{DOI}}
\vspace{0.6cm}

\noindent \normalsize{We compare the capillary levelling of a random surface perturbation on a thin polystyrene film with a theoretical study on the  two-dimensional capillary-driven thin film equation. Using atomic force microscopy, we follow the time evolution of samples prepared with different initial perturbations of the free surface.  In particular, we show that the surface profiles present long term self-similarity, and furthermore, that they converge to a universal self-similar attractor that only depends on the volume of the perturbation, consistent with the theory. Finally, we look at the convergence time for the different samples and find very good agreement with the analytical predictions.}
\vspace{0.5cm}
\end{@twocolumnfalse}]
 
\section*{Introduction}
\footnotetext{\textit{$^{a}$~Laboratoire de Physico-Chimie Th\'eorique, UMR CNRS 7083 Gulliver, ESPCI ParisTech, PSL Research University.}}
\footnotetext{\textit{$^{b}$~Department of Physics \& Astronomy and the Brockhouse Institute for Materials Research, McMaster University, Hamilton, Canada}}
\footnotetext{\textit{$^{*}$~These authors contributed equally to this work.}}
\footnotetext{\textit{$^{\ddagger}$~E-mail: elie.raphael@espci.fr }}

In the past decades, thin films have been of undeniable interest to scientific and industrial communities\cite{Blossey,Oron1997,Craster2009}. Indeed, understanding the dynamics and stability of thin films is essential to technological applications such as  nanolithography\cite{Tei11, Cho95} and the development of non-volatile memory storage devices\cite{Ouy04}. Moreover, thin films have enabled the study of the effect of confinement on polymers\cite{Gra94, For96, Si05, Bod06, Shi07, Fak08, Bae09, Rae10, Tho11, Li12}.  Several experiments have been performed in order to gain insights into the dynamics of these films. Examples are provided by the broad class of dewetting experiments\cite{Sro86, Bro90, Rei92, Rei93, Red94, Bro94, Xie98, See01b, Net04, Vil06, Bau09, Sno10, Bau12}, as well as studies on capillary levelling \cite{kerle01MAC, buck04MAC, teisseire11APL, zhu11PRL, rognin11PRE, rognin12JVS, McG12, Sal12b,  Bom13,Bau13,Backholm2014}.  Levelling experiments on thin polymer films in the vicinity of the glass transition temperature have recently given insights into the surface flow in glassy polymers\cite{Chai14}. The effect of viscoelasticity related to the polymeric nature of these films has been addressed as well\cite{Rauscher2005, Munch2006, Ben14}.

\smallskip

Thin liquid films are also of great interest to the hydrodynamics and applied mathematics community, as the viscous relaxation of a perturbed free surface is described by a nonlinear partial  differential equation that, to date, remains only partially solved. This equation is called the capillary-driven thin film equation\cite{Blossey,Oron1997,Craster2009}. Several analytical \cite{Bowen2006, Myers1998, Ben2013} and numerical\cite{Bertozzi1998} studies have allowed for a deeper understanding of its mathematical features. Recently, it was shown that the solution of the thin film equation for any sufficiently regular initial surface profile uniformly converges in time towards a universal self-similar attractor that is  given by the Green's function of the linear  capillary-driven thin film equation\cite{Ben2013}. In the terminology of Barenblatt\cite{Barenblatt1996}, this attractor corresponds to the intermediate asymptotic regime. ``Intermediate'' refers to time scales that are large enough for the system to have forgotten the initial condition, but also far enough from the generally predictable final equilibrium steady state; which, for  capillary-driven thin films is a perfectly flat surface. For thin films, the question of the convergence time to this universal attractor has not been addressed so far and is the focus of this paper.
\smallskip

Here, we report on levelling experiments on thin polystyrene films that corroborate the theoretical predictions on the convergence of the surface profiles to a universal self-similar attractor. In the first part, we recall the main results of the theoretical derivation of the intermediate asymptotic regime, and address the question of the convergence time.  In the second part, we present the experiments where we follow the time evolution of samples prepared with different random initial perturbations of the free surface. Consistent with the theory, we show that the surface profiles present long term self-similarity, and converge to a universal self-similar attractor that only depends on the volume of the  perturbation. In particular, the convergence times measured in the different samples show very good agreement with the theory. 

\section{Theory}
\label{sec:theory}

Here we recall the main theoretical results  from our previous work\cite{Ben2013}, and derive an expression for the convergence time as a function of the  volume of the perturbation.

\subsection{Levelling of a thin liquid film}

The levelling of a supported thin liquid film can be described within the lubrication approximation. Assuming incompressible viscous flow, together with a no-slip boundary condition at the substrate and a no-stress boundary condition at the free surface,  yields the so-called capillary-driven  thin film equation\cite{Blossey,Oron1997,Craster2009}:
\begin{eqnarray}
\partial_t h +\frac{\gamma}{3\eta}\, \partial_x\left(h^3\partial_x^3 h\right)&=&0 \ ,\label{TFE}
\end{eqnarray}
where $h(x,t)$ is the thickness  of the film at position $x$ and time $t$, $\gamma$ is the surface tension, and $\eta$ is the viscosity.  Equation \eqref{TFE} can be nondimensionalised  through $h=h_0 H$, $x=h_0 X$ and $t=(3\eta h_0 /\gamma) T$, where $h_0$ is the equilibrium thickness of the film infinitely far from the perturbation. This  leads to:
\begin{eqnarray}
\partial_T H + \partial_X\left(H^3\partial_X^3 H\right)&=&0 \ .\label{TFEad}
\end{eqnarray}
The height of the film can be written as $h(x,t)=h_0+\delta(x,t)$, where $\delta(x,t)$ is the perturbation that levels with the passing of time. For the case of small perturbations compared to the overall thickness of the film, Eq.~\eqref{TFEad} can be linearised by letting $H(X,T) =1+\Delta(X,T)$ where $\Delta(X,T)\ll 1$. This yields the linear thin film equation:
\begin{eqnarray}
\partial_T \Delta + \partial_X^4\Delta&=&0 \ .\label{LTFE}
\end{eqnarray}
For a given sufficiently regular initial condition $\Delta(X,0)=\Delta_0(X)$, the solution of Eq.~\eqref{LTFE} is given by:
\begin{eqnarray}
\Delta(X,T)&=&\int{\text d X'}  \, \mathcal G(X-X',T)\, \Delta_0(X') \ , \label{Sol}
\end{eqnarray}
where $\mathcal G$ is the Green's function of Eq.~(\ref{LTFE}), and reads\cite{Ben2013}:
\begin{eqnarray}
\mathcal G(X,T)&=&\frac1{2\pi}\int{\text d K} \,e^{-K^4T}e^{iKX} \ . \label{Green}
\end{eqnarray}

By 'sufficiently regular', we mean in particular that the initial perturbation of the profile is summable, with a non-zero algebraic volume, and that this perturbation vanishes when $X\rightarrow\pm\infty$.
The Green's function is obtained by taking the spacial Fourier transform of Eq.~\eqref{LTFE}. Equations~\eqref{Sol} and \eqref{Green} are central to the problem as, for a given initial condition,  they give the profile at any time.

\subsection{Universal self-similar attractor}\label{s-s}

\begin{figure}[t]
\centering
\includegraphics[width=8.6cm]{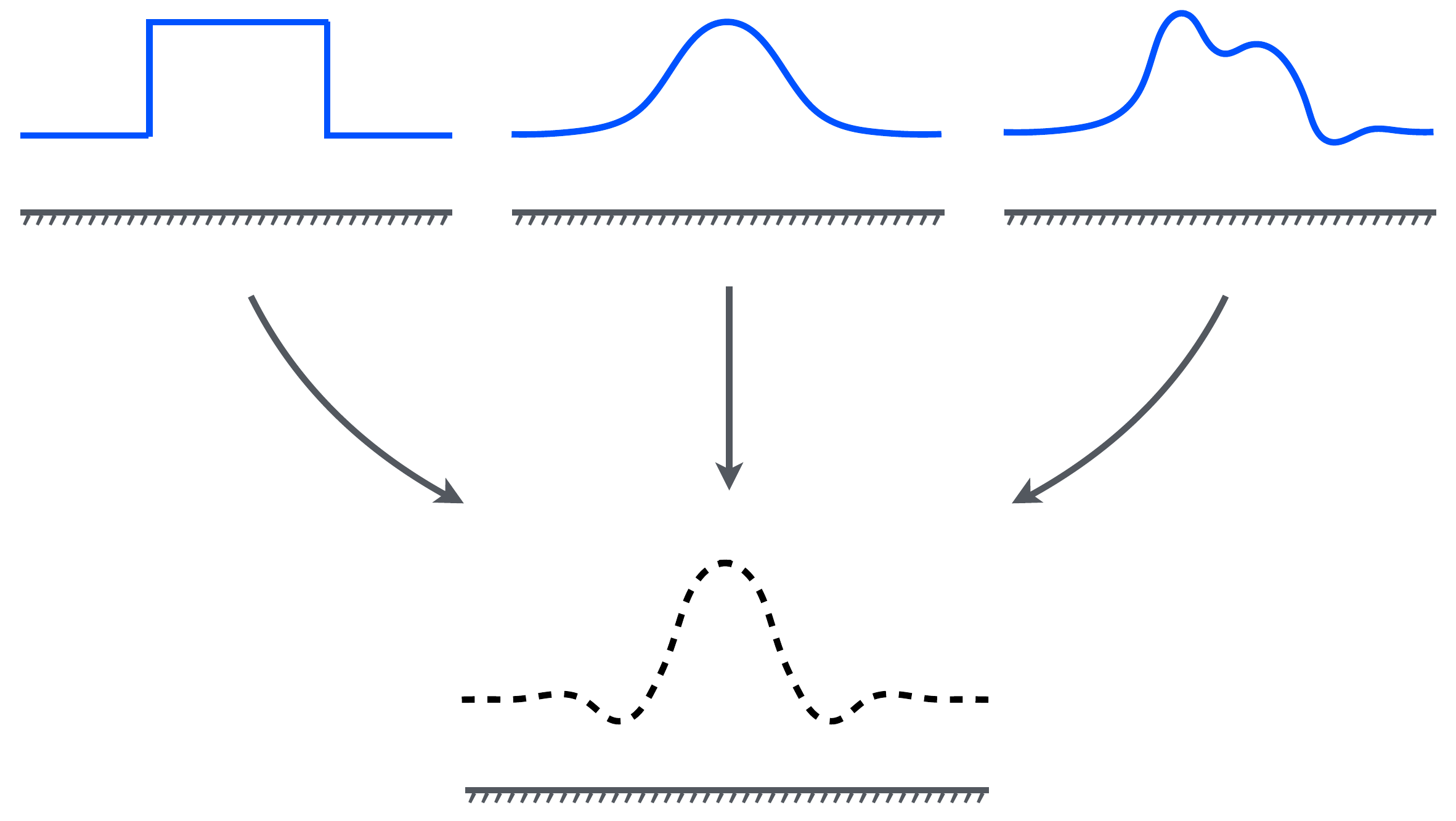}
\caption{ Schematic illustrating the convergence of any given initial profile to the universal intermediate asymptotic solution.}
\label{fig1}
\end{figure}

Guided by the mathematical structure of Eq.~\eqref{LTFE}, we introduce the self-similar change of variables: $U=XT^{-1/4}$, and $Q=KT^{1/4}$, together with $\breve \Delta(U,T)= \Delta(X,T)$. These variables, together with Eqs.~\eqref{Sol} and \eqref{Green}, yield:
\begin{eqnarray}
\breve \Delta(U,T)&=&\int{\text d X'}  \, \breve {\mathcal G}(U-X'T^{-1/4},T)\, \Delta_0(X') \ , \label{SolU}
\end{eqnarray}
where $\breve {\mathcal G}(U,T)=T^{-1/4}\phi(U)$, and:
 \begin{eqnarray}
\phi(U)&=&\frac{1}{2\pi}\int{\text d Q} \,e^{-Q^4}e^{iQU} \ . \label{Phi}
\end{eqnarray}
Note that the integral in Eq.~\eqref{Phi} can be expressed in terms of hypergeometric functions (see appendix).
The main result from our previous work \cite{Ben2013} was  that, for any given initial condition $\Delta_0(X)$ the rescaled solution  $T^{1/4} \breve {\Delta}(U,T)/\mathcal M_0$,  where  $\mathcal M_0=\int\text dX \,\Delta_0(X)\neq0$ is the the algebraic volume of the  perturbation, uniformly converges in time to $\phi(U)$  (see Fig.~\ref{fig1}): 
\begin{eqnarray}
\lim_{T\rightarrow \infty}  \frac{ T^{1/4} \breve {\Delta}(U,T)}{\mathcal M_0}&=&\phi(U)  \ . \label{Attrac}
\end{eqnarray}
According to Barenblatt's theory\cite{Barenblatt1996}, this is the intermediate asymptotic solution. The solution is universal in the sense that it does not depend on the shape of the initial condition.  Note that in the particular case of a zero volume perturbation, the attractor is given by the derivatives of the function $\phi(U)$. The question of the time needed to reach this fundamental solution is important as it quantifies how long one has to wait to forget the initial condition.

\subsection{Convergence time}\label{convergence}

In order to study the approach to the self similar attractor, we look at the surface displacement at $x=0$ as a function of time. Letting $\Delta_{\infty}(X,T)$ be the perturbation profile in the intermediate asymptotic regime, then according to Eq.~\eqref{Attrac} at $U=0$ one has:
\begin{eqnarray}
\Delta_{\infty}(0,T)&=& {\mathcal M_0\, \phi(0)}/{T^{1/4}}\ .
\end{eqnarray}
 We then define the convergence time $T_{\textrm c}$ as being the intersection of the  initial central height and the central height in the intermediate asymptotic regime: 
 \begin{eqnarray}
 \Delta_0(0)&=&\Delta_{\infty}(0,T_{\textrm c}) \ ,
 \label{e:convergence} 
 \end{eqnarray}
which leads to:
\begin{eqnarray}
    T_{\textrm c}&=& \left(\frac{ \Gamma(5/4)}{\pi} \frac{\mathcal M_0}{\Delta_0(0)} \right)^4\ .
    \label{e:master}
\end{eqnarray}
Note that the choice of origin, $x=0$, is arbitrary and will be discussed in the experimental section.

\section{Experiments}

Samples were prepared using polystyrene (PS) with weight averaged molecular weight $M_w = 31.8 \; \text{kg/mol}$ and polydispersity index $\text{PI} = 1.06$ (Polymer Source Inc.). Solutions of PS in toluene (Fisher Scientific, Optima grade) were prepared with various weight fractions, $1 < \phi < 10 \; \text{wt} \%$. Films with thickness $h_{\mathrm{Si}}$ were spincast onto clean 10 mm $\times$ 10 mm Si wafers (University Wafer) and films with thickness $h_{\mathrm{Mi}}$ were spincast onto freshly cleaved 25 mm $\times$ 25 mm mica substrates (Ted Pella Inc.).

To prepare samples with various surface geometries the following procedure was used. First, $\sim$ 10 mm $\times$ 10 mm sections of the films prepared on mica were floated onto the surface of an ultrapure water bath (18.2 M$\Omega$cm, Pall, Cascada, LS). These pieces of film were then picked up using the previously prepared films with thickness $h_{\mathrm{Si}}$ on the Si substrate. During this transfer, the floating films were intentionally folded back on themselves to create random non-uniform surface geometries. We emphasize that samples were prepared at room temperature, well below the glass transition temperature $T_{\textrm g} \approx 100 \; ^\circ \text{C}$. Two types of samples were prepared:
\begin{itemize}
\item  \textit{small perturbations}: Films with a relatively small thickness perturbation, where the linear thin film equation is expected to be valid. Such films were prepared with thicknesses $h_{\mathrm{Mi}} \ll h_{\mathrm{Si}}$ to create surface perturbations with $\max[\delta(x, 0)] /h_0  \ll  1$. We used film thickness combinations $\{h_{\mathrm{Si}}, h_{\mathrm{Mi}}\} \approx$  \{600~nm,~80~nm\} and  \{200~nm,~25~nm\}.
\item \textit{large perturbations}: Films with large thickness perturbations relative to $h_0$. Varying geometries were prepared with thicknesses $h_{\mathrm{Mi}} \approx h_{\mathrm{Si}}$ to create surface perturbations with $\max[\delta(x, 0)] /h_0 \sim 1$. Samples were prepared using film thickness combinations $\{h_{\mathrm{Si}}, h_{\mathrm{Mi}}\} \approx$  \{100~nm,~100~nm\}, \{150~nm,~150~nm\}, and \{200~nm,~200~nm\}.
\end{itemize}

The shapes of the non-uniform perturbations were not prepared by design, rather, during the preparation process many profiles are found on a single sample. Regions of interest were then located and chosen such that, while the height is varying in one direction, it is sufficiently invariant in the orthogonal horizontal direction, \textit{i.e.} $h$ can be taken to be a function of $x$ and $t$ alone. Ensuring that the profiles were invariant in one direction was crucial for the comparison to the two-dimensional theory discussed above. Having prepared non-uniform surface perturbations, a second piece of film with thickness $h_{\mathrm{Mi}}$ was floated onto a portion of the sample with thickness $h_{\mathrm{Si}}$ to create a stepped bilayer geometry, the details of which are fully explained elsewhere \cite{McG12}. Briefly, the initial height profile of such a step is well described by a Heaviside step function. When a stepped film profile is annealed above $T_{\textrm g}$ the step levels due to capillary forces. For this well defined and well studied geometry, measuring the evolution of the film height profile over time gives an in situ measurement of the capillary velocity, $\gamma/\eta$. We emphasize that each sample has \textit{both} the perturbation of interest as well as a region where there is a stepped bilayer. By obtaining the capillary velocity $\gamma/\eta$ from the bilayer portion of the sample while also probing the perturbation on the same sample, we reduce measurement error (for example due to small sample-to-sample variations in annealing temperature). The final stage in the preparation of the samples is a 1 min anneal at $130\; ^\circ \text{C}$ on a hot stage (Linkam Scientific Instruments Inc.)  to ensure that the floated films were in good contact with the substrate film and to remove any water from the system. Note that although there is some evolution of the geometry during this short initial annealing stage, as will become clear below, $t=0$ is defined after this annealing step. 

The initial film height profiles of both the surface perturbation and stepped bilayer were measured with AFM (Veeco, Caliber). In order to measure the evolution of the surface profiles, samples were annealed under ambient conditions on the hot stage at $140 \; ^\circ$C using a heating rate of $90 \; ^\circ$C/min. Above $T_{\textrm g}$, capillary forces drive the non-uniform surface geometries to level. After some time the samples were  rapidly quenched to room temperature and both the perturbation and bilayer film profiles were measured using AFM. From the AFM scans of the stepped bilayer (not shown), we use the technique described previously\cite{McG12} to extract the capillary velocity. For all samples we measure the capillary velocity $\gamma/\eta \approx 50\ \mu \text{m}/\text{min}$, which is in excellent agreement with previous measurements~\cite{Bau13, McG12}.

\section{Results and Discussions}
\subsection{Small perturbations}
\begin{figure*}[t]
\centering
\includegraphics[width=18cm]{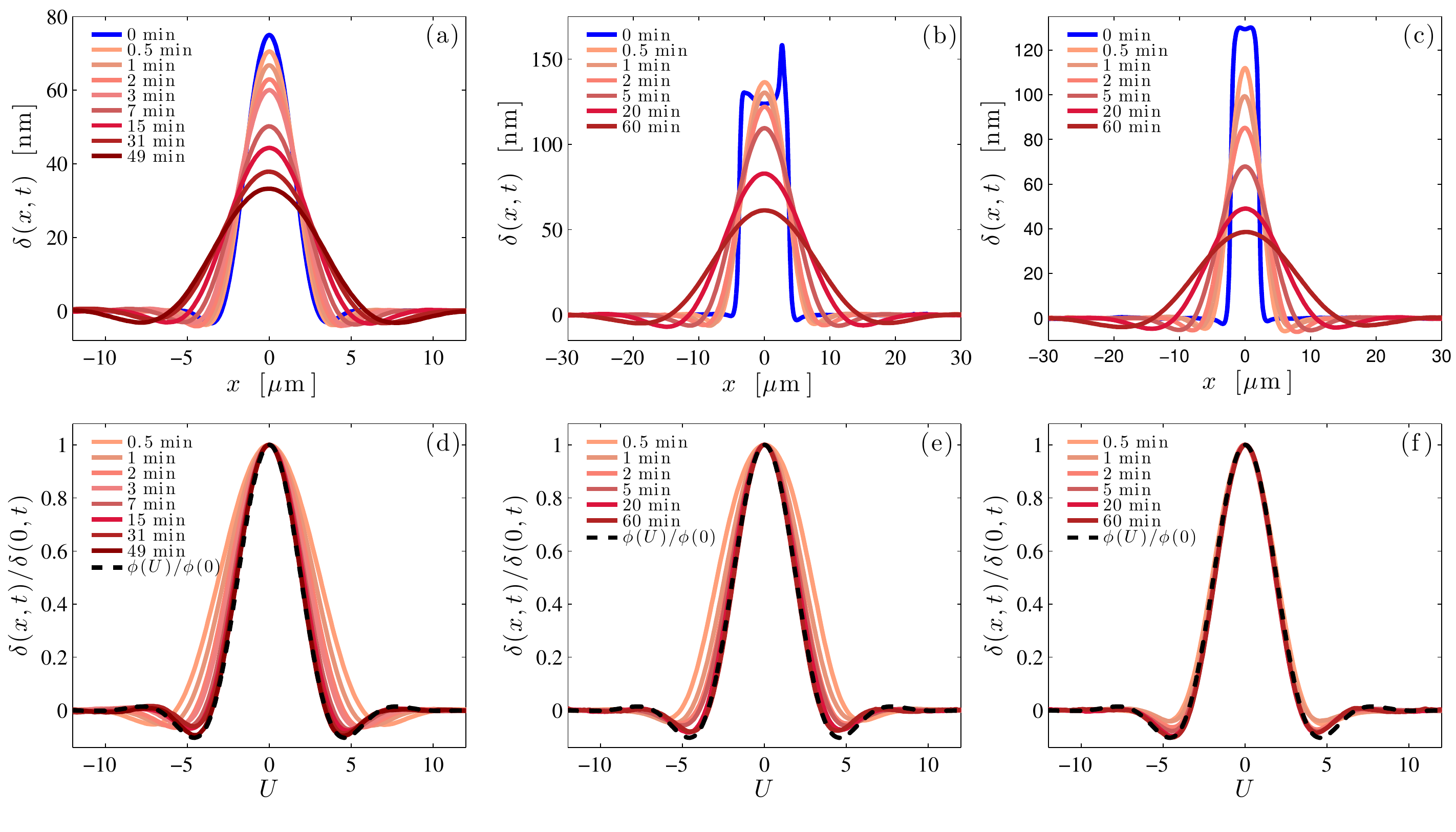}
\caption{The results of three experiments on small perturbations. The top panel shows the height of the perturbation, $\delta(x,t)= h(x,t)-h_0$,  as a function of position for annealing times $0 \leq t \leq 60 \; \text{min}$ for samples with (a) $h_0 = 221$~nm, (b) $h_0=681$~nm, and (c) $h_0=681$~nm. The bottom row shows the height of the perturbation scaled by the height at $x=0$ as a function of  $U=XT^{-1/4}=x(3\eta/h_0^3\gamma \,t)^{1/4}$. For comparison, we also plot the rescaled self-similar attractor (see Eq.~(\ref{Phi})) which is shown as a black dashed line in the bottom row. }
\label{fig2}
\end{figure*}

In Fig.~\ref{fig2}(a), (b) and (c) are shown the evolution of three examples of small perturbations, with the highest profiles corresponding to the initial $t=0$ profiles. Here, we have chosen the coordinate $x=0$ such that the volume of the perturbation for $x<0$ is equal to that of $x>0$. In the initial stages of annealing, the perturbations quickly lose any asymmetry in their shape. With additional annealing, the symmetric profiles broaden and their maximal heights decrease. Since the heights of the linear profiles are small compared to the equilibrium film thicknesses $h_0$, we expect their evolution to be governed by Eq.~(\ref{LTFE}) (the linearized thin film equation). In particular, at long times we expect the profiles to converge to the universal self-similar attractor described in Section~\ref{s-s}.

To test this prediction we plot the normalized height of the perturbation as a function of the variable $U=XT^{-1/4}=x(3\eta/h_0^3\gamma \,t)^{1/4}$, as shown in Fig.~\ref{fig2} (d), (e) and (f). We observe that at late times the profiles converge to the  rescaled self-similar attractor $\phi(U)/\phi(0)$ regardless of the initial condition, as predicted in Section~\ref{s-s}.  Here, we emphasize that since we have determined the capillary velocity in situ by measuring the evolution of a stepped bilayer geometry near the perturbation on each sample, there is no free parameter in the above rescaling and comparison to the theoretical prediction (shown as a dashed black line in Fig.~\ref{fig2} (d), (e) and (f)). Furthermore, at late times, the error between the experimentally measured profiles and the attractor is less than  $1\%$. 

\subsection{Large perturbations}
\begin{figure}[t]
\centering
\includegraphics[width=8cm]{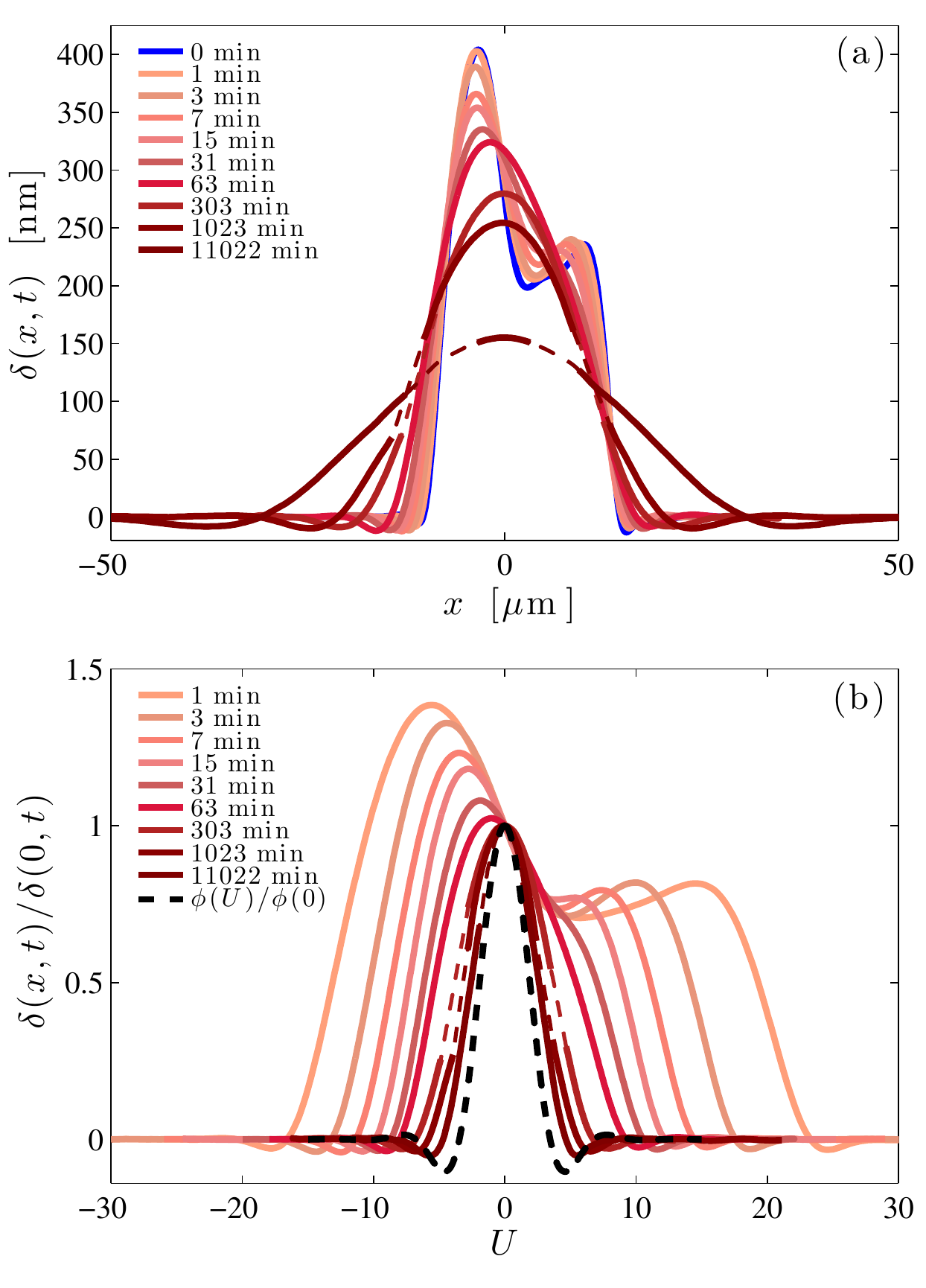}
\caption{An example data set with large perturbation. (a) Height of the perturbation as a function of position and annealing time with $h_0=216$~nm; (b) the normalized profiles. For times $t \geq 303$~min, profiles were measured using imaging ellipsometry (IE). Regions where IE is insensitive have been interpolated with quadratic splines as indicated by the dashed lines. The black dashed line corresponds to the rescaled self-similar attractor (see Eq.~(\ref{Phi})). }
\label{fig3}
\end{figure}

Measurement of the samples with large perturbations (see example in Fig.~\ref{fig3}) were more challenging because at long annealing times ($t > 100$ min) the lateral extent of the height profiles exceeds the accessible range of the AFM ($\sim 100\ \mu$m). Here, we resort to imaging ellipsometry (Accurion, EP3) to record height profiles. Imaging ellipsometry (IE) has $\sim$~nm height resolution with lateral resolution comparable to an optical microscope: $\sim \mu$m.  Thus, IE and AFM are complimentary techniques. For the example in Fig.~\ref{fig3}, data was acquired with AFM for $t \leq 63$~min, while IE was used for the three longest annealing times. With IE there is one caveat: in certain ranges of thickness there is a loss of sensitivity depending on the wavelength of laser light and the angle of incidence used (658~nm, 42-50~deg).\footnote[2]{This issue can be circumvented by varying the angle of incidence. However this was not possible for the experiments presented here because changing angle of incidence also shifts the region of interest slightly.}  For the IE data the regions where the IE was insensitive were interpolated with a quadratic spline as indicated by dashed lines to guide the eye. 

The evolution of a large perturbation is shown in Fig.~\ref{fig3}. In this case, the perturbation does not obey the condition $\delta(0,0)/h_0 \ll 1$. As can be seen in Fig.~\ref{fig3}(a), with sufficient annealing, the large perturbations become symmetric. Similar to the evolution observed for the small perturbations, once the profiles are symmetric, the maximal height $\delta(0,t)$ decreases with further annealing and the profiles broaden. 

The normalized profiles are shown in Fig.~\ref{fig3}(b). Although the perturbations are initially large, upon long enough annealing the condition $\delta(0,t) \ll h_0$ can be reached. In particular, the final state of a large perturbation is still expected to be the self-similar attractor.  For the data shown in Fig.~\ref{fig3}, even after 11022 min of annealing, the profile has not reached the condition that $\delta(0,t) \ll h_0$. While the height profiles are clearly symmetric at long times, and they are converging towards self-similarity, the final profile is not yet equivalent to the final attractor of Fig.~\ref{fig2}. The fact that the sample has not yet fully reached the self-similar attractor is simply because the starting profile was so tall, that the long annealing times required and the width of the profile (while still requiring good height resolution) place this outside our experimental window.

\subsection{Convergence Time}
\begin{figure}[t]
\centering
\includegraphics[width=8cm]{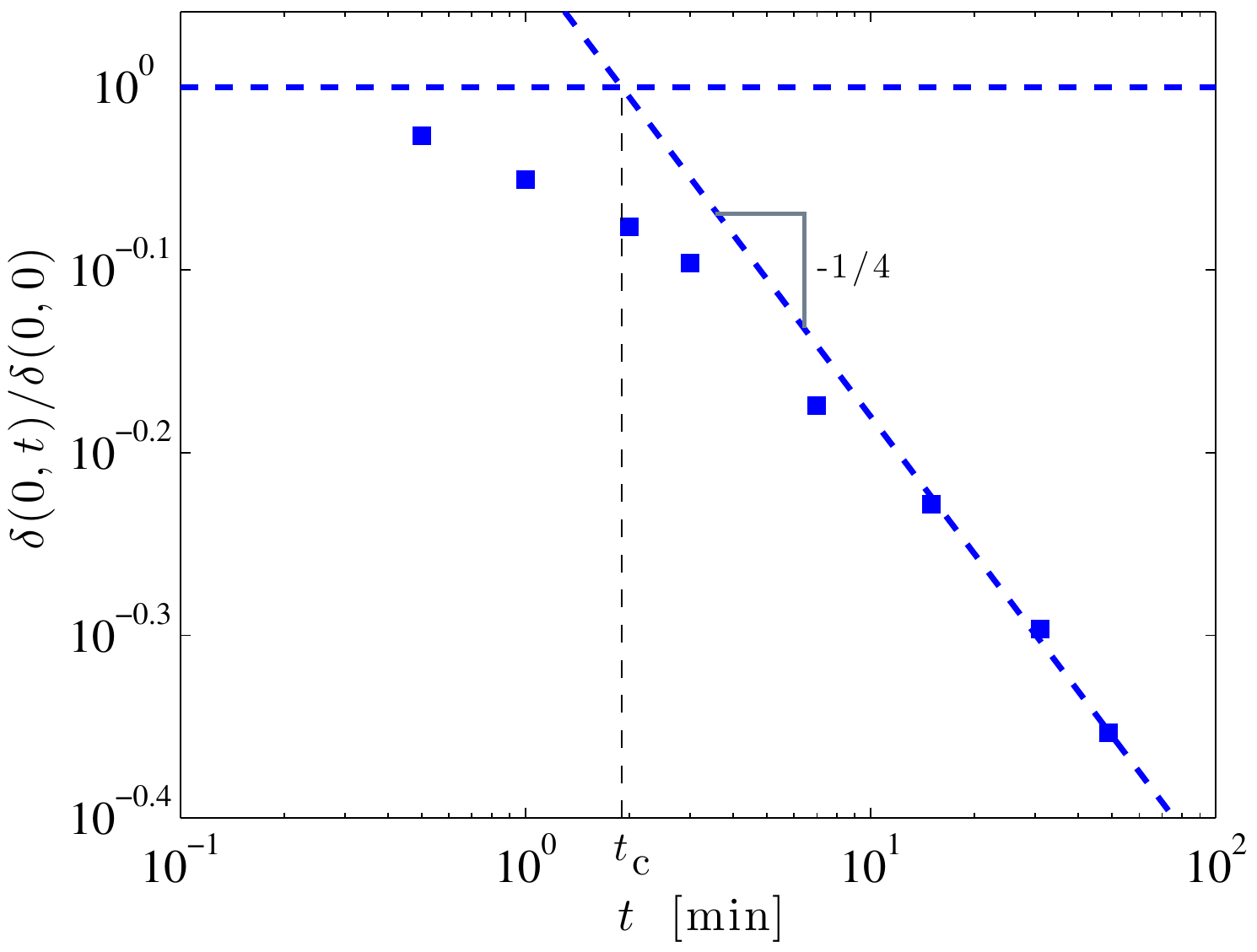}
\caption{Central height of the small perturbation shown in Fig.~\ref{fig2}(a) and (d) normalized by its initial value as a function of time. The horizontal dashed line represents the initial value. A power law of $t^{1/4}$ is fit to the late time data. In  accordance with Eq.~(\ref{e:convergence}), the convergence time is defined as the intersection of these two regimes, as indicated by the vertical dashed line.}
\label{fig4}
\end{figure}

One of the main predictions of the theory outlined in Section~\ref{convergence} is that the time taken to converge to the attractor depends on the algebraic volume of the perturbation according to Eq.~(\ref{e:master}). The convergence time is determined in accordance with Eq.~(\ref{e:convergence}) as the crossover from an initial regime, which is highly dependent on $\delta(x,0)$,  to a universal intermediate asymptotic regime. In Fig.~\ref{fig4}, we plot the normalized central height of the perturbation, $\delta(0,t)/\delta(0,0)$, for the small perturbation shown in Fig.~\ref{fig2}(a) and (d). The initial state can be characterized by the central height of the perturbation at $t=0$ and is given by the horizontal line. At late times, the maximal height of the normalized perturbation $\delta(0,t)/\delta(0,0)$ decreases in time following the $t^{-1/4}$ power law. Note that the $t^{-1/4}$ line is fit to the last three data points which correspond to the latest profiles shown in Fig.~\ref{fig2}(d). These three profiles are in excellent agreement with the calculated asymptotic profile (see black dashed line in Fig.~\ref{fig2}(d)). The crossover from the initial regime to the  intermediate asymptotic regime shown in Fig.~\ref{fig4} gives the experimentally determined convergence time, $t_{\textrm c}$. From $t_{\textrm c}$, the non-dimensionalized convergence time, $T_{\textrm c}$, can be obtained.  

\begin{figure}[t]
\centering
\includegraphics[width=9cm]{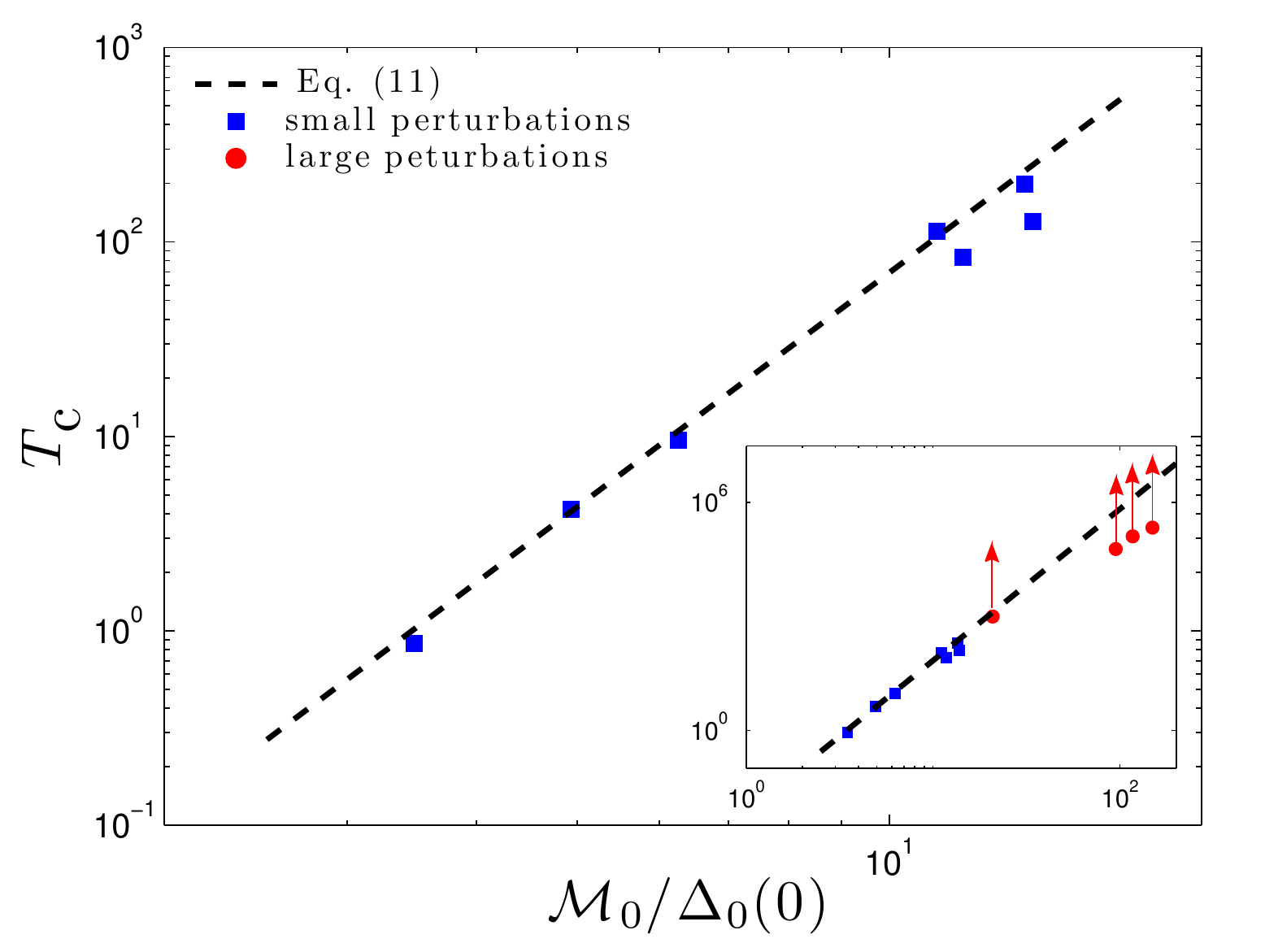}
\caption{Non-dimensionalized convergence time as a function of non-dimensionalized width. Here the square data points represent data from small perturbation samples which are in excellent agreement with the dashed black line.  The dashed black line is the theoretical prediction of Eq.~\ref{e:master}. In the inset both the small perturbation results, which have reached the self-similar regime, and the large perturbation data (circles and arrows) which are not yet self-similar are shown. The large perturbation data provides only a lower-bound for $T_{\textrm c}$, which is why that data falls below the predicted line. }
\label{fig5}
\end{figure}

The theory predicts a very clear dependence of the dimensionless convergence time, $T_{\textrm c}$, on $M_0/\Delta_0(0)$, a measure of the dimensionless width of the initial profile (see Eq. (\ref{e:master})). In Fig.~\ref{fig5} is plotted the dimensionless convergence time  obtained as in Fig.~\ref{fig4} as a function of $M_0/\Delta_0(0)$, for seven small perturbations, as well as four large perturbations. For  small perturbations, we observe excellent agreement between experiments and the theoretical prediction of Eq.~(\ref{e:master}) with no fitting parameters. We also show the convergence time for the large perturbation data (see inset of Fig.~\ref{fig5}). However, since the large perturbations have not fully reached the intermediate asymptotic regime, the $T_{\textrm c}$ one obtains by forcing a $t^{-1/4}$ power law through the latest data point corresponds to a lower bound. For this reason, the data points provided are shown with vertical arrows.

\section*{Conclusions}
We have studied, both with theory and experiment, the capillary-driven levelling of an arbitrary surface perturbation on a thin liquid film. Using atomic force microscopy and imaging ellipsometry we follow the evolution of the perturbations and compare the results to the theoretical predictions of the  two-dimensional  capillary-driven thin film equation. We have shown that regardless of the initial condition, the perturbations converge to a universal self-similar attractor that is given by the Green's function of the linear  thin film equation. Furthermore, we have shown that the time taken to converge to the attractor depends on the volume of the perturbation. We measured the convergence time for both small and large perturbations and found good agreement between theory and experiment. Specifically, the experimental results are consistent with the theory over two orders of magnitude in the dimensionless typical width of the initial profile and six orders of magnitude in dimensionless convergence time, with no free parameter.
\section*{Acknowledgements}
The financial support by NSERC of Canada and \'Ecole Normale SupŽ\'erieure of Paris, are gratefully acknowledged. The authors also thank M. Ilton, J. D. McGraw, M. Backholm, O. B\"aumchen, and H. A. Stone for fruitful discussions, as well as Etienne Rapha\"el for the cover artwork.

\section*{Appendix}

\noindent We here wish to calculate the integral in Eq.~\eqref{Phi} in terms of Hypergeometric functions.
Performing a Taylor expansion of the integrand yields:
\begin{eqnarray}
\phi(U)&=&\frac{1}{2\pi}\sum_{k=0}^\infty  \frac{(iU)^k}{k!} \left(\, \int \text dQ \,Q^k\, e^{-Q^4}    \right) \ .
\end{eqnarray} 
At this stage one can see that all terms corresponding to an odd $k=2p+1$ are null and thus that the function $f$ is real. Furthermore, changing the variables through $S=Q^4$ leads to:
\begin{eqnarray}
\phi(U)&=&\, \frac 1{4\pi} \sum_{p=0}^\infty  \frac{(iU)^{2p}}{(2p)!} \int_0^{\infty} \text dS \,S^{(1+2p)/4-1}\, e^{-S}    \ ,
\end{eqnarray} 
where we recognise a $\Gamma$ function:
\begin{eqnarray}
\phi(U) &=& \frac1{4\pi} \sum_{p=0}^\infty  \frac{(iU)^{2p}}{(2p)!} \, \Gamma\left( \frac{1+2p}{4}\right) \ .
\end{eqnarray} 
Then, separating the sum over $p$ in even $p=2m$ and odd $p=2m+1$ terms yields:
\begin{eqnarray}
\phi(U)&=& \frac1{4\pi} \sum_{m=0}^\infty  \frac{U^{4m}}{(4m)!} \, \Gamma\left( m+\frac{1}{4}\right) \nonumber \\
&& -\,\,\,\frac{U^2}{4\pi} \sum_{m=0}^\infty  \frac{U^{4m}}{(4m+2)!} \, \Gamma\left(m+ \frac{3}{4}\right)  \ .
\end{eqnarray} 
Developing the $\Gamma$ functions in terms of Pochhammer rising factorials $\Gamma(m+\alpha)=\Gamma(\alpha) (\alpha)_m$ where $(\alpha)_m=\alpha\times (\alpha+1)\times ... \times(\alpha+m-1)$,
and using the relation  $\Gamma(\alpha+1)=\alpha\Gamma(\alpha)$ yields:
\begin{eqnarray}
\phi(U)&=&\frac{1}{\pi}\,\Gamma\left(  \frac54\right)\sum_{m=0}^\infty  \frac{U^{4m}}{(4m)!} \, \left(\frac{1}{4}\right)_m \nonumber \\
&&  -\,\, \frac {U^2}{4\pi} \,\Gamma\left( \frac34  \right) \sum_{m=0}^\infty  \frac{U^{4m}}{(4m+2)!} \, \left(\frac{3}{4}\right)_m  \ .
\end{eqnarray} 
Developing the factorials and rising factorials
and proving by mathematical induction that:
\begin{eqnarray}
 4^{3m}\, \frac{ 1\times 5\times ... \times (1+4(m-1))  }{4m\times (4m-1)\times ...\times 1}=\frac{1}{m!}\,  \frac 1{\left(\frac{1}{2}\right)_m\left(\frac{3}{4}\right)_m}   \ ,
\end{eqnarray} 
and that:
\begin{eqnarray}
4^{3m}\, \frac{ 3\times 7\times ...\times (3+4(m-1)) }{(4m+2)(4m+1) \times ... \times 1} =\frac12\, \frac{1}{m!} \, \frac 1{\left(\frac{3}{2}\right)_m\left(\frac{5}{4}\right)_m}   \ ,
\end{eqnarray} 
finally leads to:
\begin{eqnarray}
\phi(U) &=& \frac{1}{\pi}\,\Gamma\left(\frac{5}{4}\right)\ _0H_{2}\left[\left\{ \frac12,\frac34\right\},\left(\frac{U}{4}\right)^4\right]  \nonumber \\
&&-\,\,\frac{U^2}{8\pi} \,\Gamma\left(\frac{3}{4}\right)\ _0H_{2}\left[\left\{ \frac54,\frac32\right\},\left(\frac{U}{4}\right)^4\right]\ .
\label{HPG}
\end{eqnarray}
where the $(0,2)$-hypergeometric function is defined as \cite{Abramowitz1965,Gradshteyn1965}: 
\begin{eqnarray}
_0H_{2}\left(\left\{ a,b\right\},w\right)&=&\sum_{m=0}^{\infty} \frac{1}{(a)_m(b)_m}\,\frac{w^m}{m!}\ .
\end{eqnarray}

\balance
\footnotesize{
\providecommand*{\mcitethebibliography}{\thebibliography}
\csname @ifundefined\endcsname{endmcitethebibliography}
{\let\endmcitethebibliography\endthebibliography}{}

}

\end{document}